\documentclass[preprint,5p,times,twocolumn]{elsarticle}
\usepackage{mathtools}
\usepackage[english]{babel}
 \usepackage{lineno}

\usepackage{float}
\newfloat{algorithm}{t}{lop}

 \usepackage{algorithm}
\usepackage {algpseudocode}
 \usepackage{graphicx}
\usepackage{wrapfig}
\graphicspath{ {images/} }

 \usepackage{hyperref}
 \hypersetup{
  colorlinks,
  citecolor=Green,
  linkcolor=Green,
  urlcolor=Green,
  pdftitle =Bidirectional Conditional Insertion Sort  algorithm; An efficient progress on the classical insertion sort,
  pdfauthor=Adnan S. Mohammed,
  pdfsubject =Computers algorithms
  }

\journal{Future Generation Computer Systems}

\begin{document}
 
\begin{frontmatter}

\title{Bidirectional Conditional Insertion Sort algorithm; 
An efficient progress on the classical insertion sort}

\author[mymainaddress]{Adnan Saher Mohammed\corref{mycorrespondingauthor}}
 \cortext[mycorrespondingauthor]{Corresponding author}
 \ead {adnshr@gmail.com}
 
\author[mysecondaryaddress]{\c{S}ahin Emrah Amrahov}

 \ead {emrah@eng.ankara.edu.tr}
\author[Fatihdaddress]{Fatih V. \c{C}elebi}
 \ead {fvcelebi@ybu.edu.tr }

\address[mymainaddress]{Y{\i}ld{\i}r{\i}m Beyaz{\i}t University, Graduate School of Natural Sciences, Computer Engineering Dept., Ankara, Turkey}
\address[mysecondaryaddress]{ Ankara University,	Faculty of  Engineering, Computer Engineering Dept., Ankara, Turkey}
\address[Fatihdaddress]{Y{\i}ld{\i}r{\i}m Beyaz{\i}t University, Faculty of Engineering and Natural Sciences, Computer Engineering Dept., Ankara, Turkey}

\begin{abstract}
In this paper, we proposed a new efficient sorting algorithm based on insertion sort concept. The proposed algorithm called Bidirectional Conditional Insertion Sort (BCIS). It is in-place sorting algorithm and it has remarkably efficient average case time complexity when compared with classical insertion sort (IS). By comparing our new proposed algorithm with the Quicksort algorithm, BCIS indicated faster average case time for relatively small size arrays up to 1500 elements. Furthermore, BCIS was observed to be faster than Quicksort within high rate of duplicated elements  even for large size array.
\end{abstract}


\end{frontmatter}

\section{Introduction}\label{section:1}
   \par Algorithms have an important role in developing process of computer science and mathematics. Sorting is a fundamental process in computer science which is commonly used for canonicalizing data. In addition to the main job of sorting algorithms,  many algorithms use different techniques to sort lists as a prerequisite step to reduce their execution time \cite{AbuDalhoum2012}.  The idea behind using sorting algorithms by other algorithm is commonly known as reduction process. A reduction is a method for transforming one problem to another easier than the first problem \cite{Sedgewick2011a}. Consequently, the need for developing efficient sorting algorithms that invest the remarkable development in computer architecture has  increased.  \par
 Sorting is generally considered  to be the procedure of repositioning a known set of objects in ascending or descending order according to  specified key values belong to these objects. Sorting is guaranteed to finish in finite sequence of steps\cite{Chhatwani2014}. 
\par Among a large number of sorting algorithms, the choice of which is the best for an application depends on several factors like size, data type and the distribution of the elements in a data set. Additionally, there are several dynamic influences on the performance of the sorting algorithm which it can be briefed as the number of comparisons (for comparison sorting), number of swaps (for in-place sorting),memory usage and recursion \cite{Alnihoud2010}. \par
 Generally, the performance of algorithms measured by the standard Big $ O(n)$ notation which is used to describe the complexity of an algorithm. Commonly, sorting algorithms has been classified into two groups according to their time complexity. The first group is $O(n^2)$ which contains the insertion sort, selection sort, bubble sort… etc.  The second group is  $O(n\log{}n)$, which is faster than the first group, includes Quicksort ,merge sort and heap sort \cite{EshanKapurParveenKumar2012}. The insertion sort algorithm can be considered as one of the best algorithms in its family ($O(n^2)$ group ) due to its performance, stable algorithm  ,in-place, and simplicity \cite{Srivastava2009}. Moreover, it is the fastest algorithm for small size array up to 28-30 elements compared to the Quicksort algorithm. That is why it has been used in conjugate with Quicksort \cite{Sedgewick1977,Sedgewick1978,wild2015average,Bentley1993} . 

\par Several improvements on major sorting algorithms have been  presented in the literature  \cite {5635119,Codingunit,Samanta2009}. 
Chern and Hwang \cite{Chern2001} give an analysis of the transitional behaviors of the average cost from insertion sort to quicksort with median-of-three. Fouz et al \cite{Fouz2011} provide a smoothed analysis of Hoare's algorithm who has found the quicksort.
Recently, we meet some investigations of the dual-pivot quicksort which is the modification of the classical quicksort algorithm.  In the partitioning step of the dual-pivot quicksort two pivots are used to split the sequence into three segments recursively. This can be done in different ways. Most efficient algorithm for the selection of the dual-pivot is developed due to Yaroslavskiy question \cite{yaroslavskiy2010question}. Nebel, Wild and Martinez \cite{Nebel2015} explain  the success of Yaroslavskiy's new dual-pivot Quicksort algorithm  in practice.  Wild and Nebel \cite{Wild2013} analyze this algorithm and show that on average it uses $1.9n\, ln\,n + O(n)$ comparisons to sort an input of size n, beating standard quicksort, which uses $2n\,ln\,n + O(n)$ comparisons. Aum\"{u}ller and Dietzfelbinger \cite{Aumuller2016} propose a model that includes all dual-pivot algorithms, provide a unified analysis, and identify new dual-pivot algorithms for the minimization of the average number of key comparisons among all possible algorithms. 
This minimum is $1.8n \, ln\,n  + O(n)$. Fredman \cite{Fredman2014137} presents a new and very simple argument for bounding the expected running time of Quicksort algorithm. Hadjicostas and Lakshmanan \cite{Hadjicostas2011} analyze the recursive merge sort algorithm and quantify the deviation of the output from the correct sorted order if the outcomes of one or more comparisons are in error. Bindjeme and Fill \cite{Bindjeme2012} obtain an exact formula for the L2-distance of the (normalized) number of comparisons of Quicksort under the uniform model to its limit.  Neininger \cite{Neininger2015} proves a central limit theorem for the error and obtain the asymptotics of the $L_3-distance$.  Fuchs \cite{Fuchs2015} uses the moment transfer approach to re-prove Neininger's result and obtains the asymptotics of the $L_p -distance$ for all $ 1\leq p <\infty$.  

Grabowski and Strzalka \cite{grabowski2006dynamic} investigate the dynamic behavior of simple insertion sort algorithm and the impact of long-term dependencies in data structure on sort efficiency. Biernacki and Jacques \cite{Biernacki2013} propose a generative model for rank data based on insertion sort algorithm.
The work that presented  in \cite {Bender2006} is called library sort or gapped insertion sort  which is trading-off between the extra space used and the insertion time, so it is not in-place sorting algorithm. The enhanced insertion sort algorithm that presented in \cite{Sodhi2013} is use approach similar to binary insertion sort in \cite{preiss2008data}, whereas both algorithms reduced the number of comparisons and kept the number of assignments (shifting operations) equal to that in standard insertion sort $O(n^2)$. Bidirectional insertion sort approaches presented in \cite{Chhatwani2014,Dutta2013}. They try to make the list semi sorted in Pre-processing step by swapping the elements at analogous positions (position 1 with n, position 2 with (n-1) and so on). Then they apply the standard insertion sort on the whole list. The main goal of this work is only to improve  worst case performance of IS  \cite{Dutta2013} . On other hand, authors in\cite{Srivastava2009}    presented a bidirectional insertion sort, firstly exchange elements using the same way  in  \cite{Chhatwani2014,Dutta2013} , then starts from the middle of the array and inserts elements from the left and the right side to the sorted portion of the main array. This method improves the performance of the algorithm to be efficient for small arrays typically of size lying from 10-50 elements \cite{Srivastava2009} . Finally, the main idea of the work that presented in \cite{Min2010}, is based on inserting the first two elements of the unordered part  into the ordered part during each iteration. This idea earned slightly time efficient but the complexity of the algorithm still $ O(n^2)$ \cite{Min2010} . However, all the cited previous works have shown a good enhancement in insertion sort algorithm either in worst case, in large array size or in very small array size. In spite of this enhancement, a Quicksort algorithm indicates faster results even for relatively small size array.
 \par In this paper, a developed in-place unstable algorithm is presented that shows fast performance in both relatively small size array and for high rate duplicated elements array. The proposed algorithm Bidirectional Conditional Insertion Sort (BCIS) is well analyzed for best, worst and average cases. Then it is compared with well-known algorithms  which are classical  Insertion Sort (IS) and Quicksort. Generally, BCIS has average time complexity very close to $O (n^{1.5})$ for normally or uniformly distributed data. In other word, BCIS has faster average case than IS for both relatively small and large size array. Additionally, when it compared with Quicksort, the experimental results for BCIS indicates less time complexity up to 70\% -10\%  within the data size range of 32-1500. Besides, our BCIS illustrates faster performance in high rate duplicated elements array compared to the Quicksort even for large size arrays. Up to 10\%-50\%  is achieved within the range of elements of 28-more than 3000000. The other pros of BCIS that it can sort equal elements array or remain equal part of an array in $ O(n)$ .
\par This paper is organized as follows: \hyperref [section:2]{section-\ref*{section:2}} presents the proposed algorithm and pseudo code,  \hyperref [section:3]{section-\ref*{section:3}} executes the proposed algorithm on a simple example array, \hyperref [section:4]{section-\ref*{section:4}}  illustrates the detailed complexity analysis of the algorithm, \hyperref [section:5]{section- \ref*{section:5}}  discusses the obtained empirical results and compares them with other well-known algorithms,  \hyperref [section:6]{section-\ref*{section:6}}  provides conclusions. Finally, you will find the important references. 

\section{The proposed algorithm BCIS}\label{section:2}
The classical insertion sort explained in \cite{Min2010,Khairullah2013,Nenwani2014} has one sorted part in the array located either on left or right side. For each iteration, IS inserts only one item from unsorted part into proper place among elements in the sorted part. This process repeated until all the elements sorted.  \par
Our proposed algorithm minimizes the shifting operations caused by insertion processes using new technique. This new technique supposes that there are two sorted parts located at the left and the right side of the array whereas the unsorted part located between these two sorted parts. If the algorithm sorts ascendingly, the small elements should be inserted into the left part and the large elements should be inserted into the right part. Logically, when the algorithm sorts in descending order, insertion operations will be in reverse direction. This is the idea behind the word ‘bidirectional’ in the name of the algorithm. 
 \par
Unlike classical insertion sort, insertion items into two sorted parts helped BCIS to be cost effective in terms of memory read/write operations. That benefit happened because the length of the sorted part in IS is distributed to the two sorted parts in BCIS. The other advantage of BCIS algorithm over classical insertion sort is the ability to insert more than one item in their final correct positions in one sort trip (internal loop iteration).

 Additionally, the inserted items will not suffer from shifting operations in later sort trips. Alongside, insertion into both sorted sides can be run in parallel in order to increase the algorithm performance (parallel work is out of scope of this paper). 
 \par In case of ascending sort, BCIS initially assumes that the most left item at $array[1]$ is the left comparator (LC) where is the left sorted part begin. Then inserts each element into the left sorted part if that element less than or equal to the LC. Correspondingly, the algorithm assumes the right most item at $array[n]$  is the right comparator (RC) which must be greater than LC. Then BCIS inserts each element greater than or equal to the RC into the right sorted part. However, the elements that have values between LC and RC are left in their positions during the whole sort trip. This conditional insertion operation is repeated until all elements inserted in their correct positions.
 \par If the LC and RC already in their correct position, there are no insertion operations occur during the whole sort trip. Hence, the algorithm at least places two items in their final correct position for each iteration.   

\par In the  pseudo code (part 1\& 2), the BCIS algorithm is presented in a format  uses functions to increase the clarity and traceability of the algorithm. However, in statements (\ref{s1} \& \ref{s2}) the algorithm initially sets two indexes, SL for the sorted left part and SR for the sorted right part to indicate on the most left item and the most right item respectively.
 \renewcommand{\thealgorithm}{}
\algnewcommand\algorithmicto{\textbf{to}}
\algrenewtext{For}[3]%
{\algorithmicfor\ $#1 \gets #2$ \algorithmicto\ $#3$ \algorithmicdo}
\begin{algorithm}[!ht]
\caption {BCIS Part 1 (Main Body)}
\begin{algorithmic}[1]
 \State $SL\leftarrow  left$ \label{s1}
  \State $SR\leftarrow  right$ \label{s2}
 \While {$SL<SR$} \label{s3}
  \State SWAP$(array ,SR ,SL+\frac{(SR-SL)}{2}) $\label{s4}
\If {$array[SL] $ = $array[SR]$} \label{s5}
\If{ISEQUAL($array,SL,SR$)=-1}  \label{s6}
\State return \label{s7}
\EndIf
\EndIf      
\If {array[SL] $>$ array[SR]} \label{s10}
\State SWAP (array, SL , SR)
\EndIf
\If {$ (SR-SL)\geq100$} \label{s13}
\For {i }{SL+1}{(SR-SL)^{0.5}}\label{s14}
\If {array[SR] $<$ array[i]}
\State  SWAP (array, SR, i)
\ElsIf  {array[SL] $>$ array[i]}
\State  SWAP (array, SL, i)
\EndIf  
\EndFor \label{s20}
\Else 
\State  $ i\leftarrow SL+1$ \label{s22}
\EndIf
\State $LC \gets array[SL]$
\State $RC \gets array[SR]$\label{s25}
 \While {$i<SR$} \label{s26}
\State $CurrItem \leftarrow array[i]$
\If{$CurrItem \ge RC$}
\State $array [i] \gets  array[SR-1] $
\State $INSRIGHT (array, CurrItem, SR, right) $
\State $SR \gets  SR-1 $                

\ElsIf{$CurrItem \le LC$}
\State $array [i] \gets  array[SL+1] $
\State $INSLEFT (array, CurrItem, SL, left) $
\State $SL \gets  SL+1 $   
\State $i \gets i+1$
\Else 
\State $i \gets i+1$
\EndIf
\EndWhile
\State $SL\leftarrow SL+1$
\State $ SR\leftarrow SR-1$
\EndWhile
\algstore{bkbreak}
\end{algorithmic}
\end{algorithm}

\begin{algorithm} [h]
\caption{BCIS Part 2 (Functions)}
\begin{algorithmic}[h]
\algrestore{bkbreak}
   \Function {ISEQUAL}{$array,SL,SR$}
\For {k}{SL+1 } { SR-1}
\If {$array[k] !=  array[SL]$}
\State   $SWAP(array, k , SL )$
\State  return  k
\EndIf
\EndFor
 \State$return -1 $  \\ 
  \Comment End the algorithm because all scanned items are equal
\EndFunction
 \Function {InsRight}{$array, CurrItem , SR , right$}
\State $ j \gets SR$
\While {$ j\le right$ {\bf and } $CurrItem > array[j]$} 
 \State $  array[j-1] \gets array[j]$
\State $ j \gets j+1$
\EndWhile
 \State $Array[j-1]\gets CurrItem$
\EndFunction

 \Function {InsLeft}{$array, CurrItem , SL , left$}
 \State $ j \gets SL$
\While {$ j\ge left$ {\bf and } $CurrItem <array[j]$} 
 \State $ array[j+1] \gets array[j]$
\State $ j \gets j-1$
\EndWhile
 \State $Array[j+1]\gets CurrItem$
\EndFunction
\Function {SWAP}{array,i,j}
\State $Temp \gets array[i]$
\State $array[i] \gets array[j]$
\State $array[j] \gets Temp$
\EndFunction
\end{algorithmic}
\end{algorithm}
 The main loop starts at statements(\ref{s3}) and stops when the left sorted part index (SL) reaches the right sorted part index (SR). 
\par
The selection of  LC and RC is processed by the statements (\ref{s4}-\ref{s25}). In order to ensure the correctness of the insertion operations  LC must be less than RC, this condition processed in statement (\ref{s5}). In case of LC equal to RC, the statement (\ref{s6}), using “ISEQUAL” function, tries to find an item not equal to LC and replace it with LC. Otherwise, (if not found) all remaining elements in the unsorted part are equal. Thus, the algorithm should terminate at the statement (\ref{s7}). Furthermore, this technique allows equal elements array to sort in only $O(n)$ time complexity. 
Statements (\ref{s4} \& \ref{s13} -- \ref{s20})  do not have an effect on the correctness of the algorithm, these statements are added to enhance the performance of the algorithm. The advantage of these techniques will be discussed in the analysis section (\hyperref[section:4]{section-\ref*{section:4}}).
 \par
The \textbf{while} statement in (\ref{s26}) is the beginning of the sort trip, as mentioned previously, conditional insertions occur inside this loop depending on the value of current item $(CurrItem)$ in comparison with the values of LC and RC.  Insertion operations are implemented by calling the functions “INSRIGHT” and “INSLEFT”. 

\section{ Example}\label{section:3}
\begin{figure}[ht]
\includegraphics [width=1\linewidth]{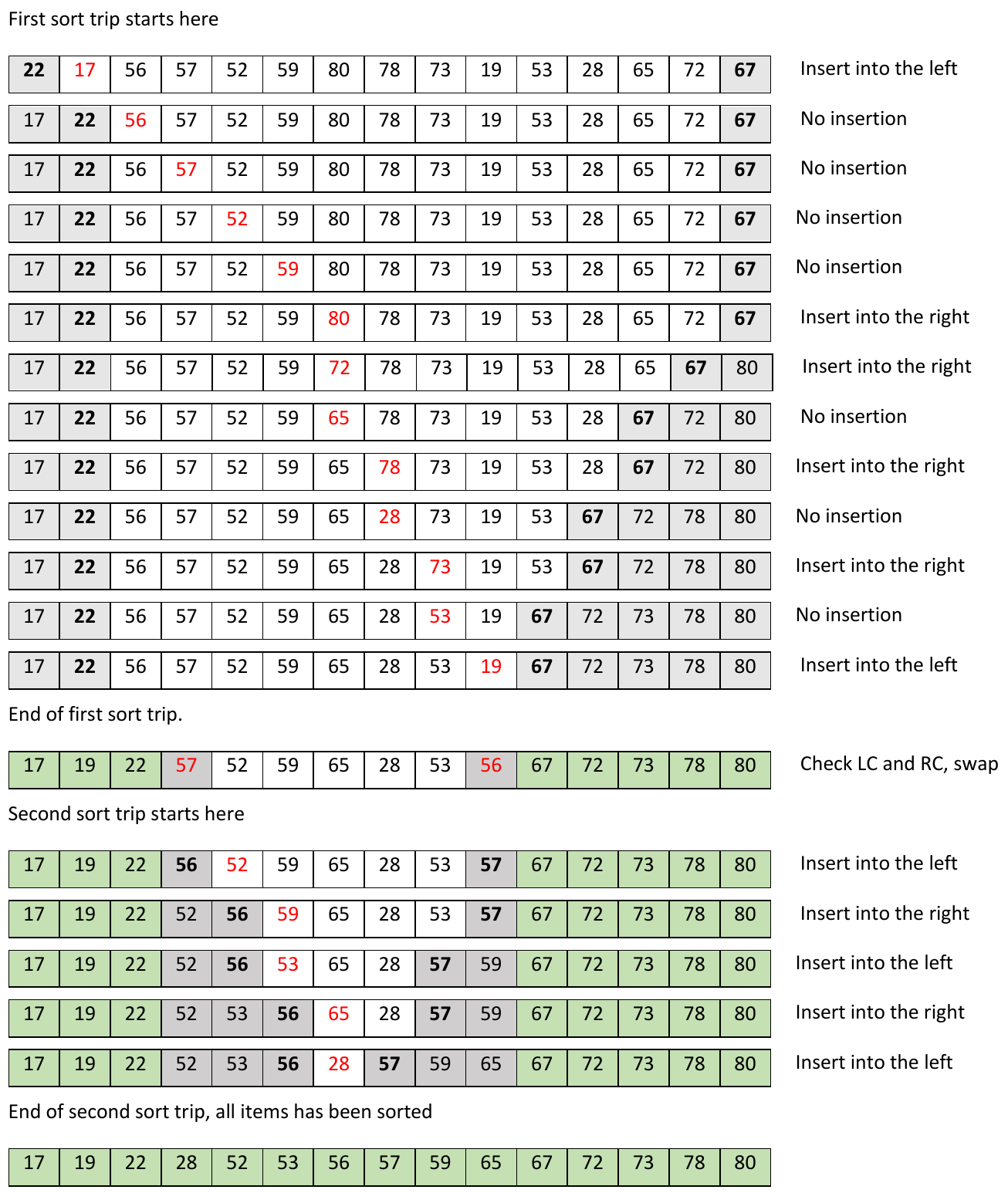}
 \caption{BCIS Example}
  \label{fig:BCISexample}
\end{figure}

 The behavior of the proposed algorithm on an array of 15 elements generated randomly by computer is explained in Figure(\ref{fig:BCISexample}). In order to increase the simplicity of this example, we assumed the statements (\ref{s4} \& \ref{s13}-\ref{s20}) do not exist in the algorithm. For all examples in this paper we assumed as follows: Items in red color mean these items are currently in process. Bolded items represent LC and RC for current sort trip. Gray background means the position of these items may change during the current sort trip. Finally, items with green background mean these items are in their final correct positions.
\section{Analysis of The Proposed Algorithm} \label{section:4}
 The complexity of the proposed algorithm mainly depends on the complexity of insertion functions which is in turn depends on the number of inserted elements in each function during each sorting trip. To explain how the performance of BCIS depends on the number of inserted element per each sort trip, several assumptions are presented which revealed theoretical analysis very close to experiential results that we obtained.
\par In order to simplify the analysis, we will concentrate on the main parts of the algorithm. Assume that during each sort trip (k) elements are inserted into both sides, each side get $ k/2$. Whereas insertion functions work exactly like standard insertion sort. Consequently, time complexity of each sort trip equal to the sum of the left and right insertion function cost which is equal to $ T_{is} (k/2)$  for each function, in addition to the cost of  scanning of the remaining elements (not inserted elements). We can express this idea as follows :- 
 \begin {align*}
   T(n)&= T_{is} (\frac{k}{2})+T_{is}(\frac{k}{2})+2(n-k)\\
          & +T_{is} (\frac{k}{2})+T_{is}(\frac{k}{2})+2(n-2k)\\ 
  \end {align*}
 \begin {align*}
         &+T_{is}(\frac{k}{2})+T_{is}(\frac{k}{2})+2(n-3k) \\
         &+\dots+T_{is}(\frac{k}{2})+T_{is}(\frac{k}{2})+2(n-ik)
   \end {align*}
BCIS stops when $ n-ik=0 \implies i=\frac{n}{k}$\\
\begin {align}
T(n)&=\frac{n}{k} \left[ T_{is} (\frac{k}{2})+T_{is} (\frac{k}{2}) \right]+\sum _{i=1}^{\frac{n}{k}}(n-ik) \label{Eq1}\\  
       &=\frac{n}{k} \left[ T_{is}(\frac{k}{2})+T_{is} (\frac{k}{2})\right]+\frac{n^2}{k}-n \notag \\ 
       &=\frac{n}{k}  \left[ T_{is} (\frac{k}{2})+T_{is} (\frac{k}{2})+n\right]-n   \label{geneq}
\end {align}
\par Equation (\ref {geneq}) represents a general form of growth function, it shows that the complexity of the proposed algorithm mainly depends on the value of k and the complexity of insertion functions.  
\subsection {Average Case Analysis}
The average case of classical insertion sort $T_{is}(n)$ that appeared in equation  (\ref {geneq})  has been well analyzed in terms of comparisons in \cite{Rosen2012,McQuain2009} and  in \cite{McQuain2009} for assignments. However, authors of the cited works  presented the following equations which represent the average case analysis for classical insertion sort for comparisons and assignments respectively.
\begin {align}
T_{isc }(n)=\frac  {n^2}{4}+   \frac {3 n}{4}-1 \label{Tisc}\\
T_{isa}(n)=  \frac{n^2}{4}+  \frac{7 n}{4}+3  \label{Tisa}
\end {align}
\par The  equations (\ref{Tisc} \& \ref{Tisa}) show that the insertion sort has approximately equal number of comparisons $(T_{isc }(n))$ and assignments $( T_{isa}(n))$. However, for BCIS, it is assumed that in each sort trip k elements are inserted into both side. Therefore, the main \textbf{while}  loop executes $n/k$ times that represent the number of sort trips.  Suppose each insertion function get $k/2$ elements where  $2\le k \le n$. Since both insertion functions (INSLEFT, INSRIGHT) exactly work as a standard insertion sort, so the average case for each function during each sort trip is . 
\begin{eqnarray}
 Comp.\# /SortTrip/Function &=&T_{isc}(\frac{k}{2}) \notag\\
                                       &=&\frac{k^2}  {16}+ \frac{3k}{8}-1 \label{compno}
\end{eqnarray}
\par BCIS performs one extra assignment operation to move the element that neighbored to the sorted section before calling each insertion function. Considering this cost we obtained as follows: -    
\begin{eqnarray}
Assig.\#/SortTrip/Function &=&T_{isa}(\frac{k}{2})+1\notag \\
                           &=&\frac{k^2}  {16}+  \frac{7k}{8}+4 \label{assigno}
\end{eqnarray}
\par In order to compute  BCIS comparisons complexity, we substituted equation (\ref{compno})  in equation (\ref {geneq}) and we obtained   as follows:-
\begin {align}
T_{c} (n)= \frac {n}{k} \left[ \frac{k^2}{8}+\frac{ 3k}{4}-2+n \right] -n  \label{Tc}
\end {align}
\par Equation (\ref{Tc}) shows that when $ k$ gets small value the algorithm performance goes close to $O(n^2)$. For $k=2$ the growth function is shown below.
\begin{align}
T_{c}(n)&= \frac{n}{2 } \left [\frac{4}{8}+\frac{3}{2}-2+n\right]- n  \notag \\
              &=\frac{n^2}{2} - n 
\end{align}
\par When k gets large value also the complexity of BCIS goes close to $O(n^2)$. For k=n the complexity is:-
\begin{align}
T_{c}(n)&= \frac{n}{n} \left [\frac{n^2}{8}+\frac{3n}{4}-2+n\right]- n  \notag \\
              &=\frac{n^2}{8} +\frac{3n}{4}- 2
\end{align}
\par Hence, the approximate best performance of the average case for BCIS that could be obtained when $ k =n^{0.5}$  as follows:-
\begin{align}
T_c (n) &= \frac {n}{n^{0.5}}   \left[ \frac {n}{8}+ \frac{3n^{0.5}}{4}-2 +n\right]-n\notag \\
            &=\frac{9 n^{1.5}}{8}-\frac{n}{4}-2n^{0.5}\label{TcBCIS}
\end{align}
\par Equation(\ref{TcBCIS}) computes the number of comparisons of BCIS when runs in the best performance of the average case. On other hand, to compute the the number of assignments for BCIS in case of best performance of average case. Since assignments operations occur only in insertions functions, equation (\ref{assigno}) is  multiplied by two  because there are two insertion functions, then the result is multiplied by the number of sort trip $\frac{n}{k}$ . When $k=n^{0.5}$ we got as follows:
\begin{align}
T_a (n) &= \frac {n}{n^{0.5}}   \left[ \frac {n}{8}+ \frac{7n^{0.5}}{4}+8\right] \notag \\
            &=\frac{ n^{1.5}}{8}+\frac{7n}{4}+8n^{0.5}\label{TaBCIS}
\end{align}
\par The comparison of equation (\ref{TcBCIS}) with equation (\ref{TaBCIS}) proves that the number of assignments less than the number of comparisons in BCIS. As we mentioned previously in equations(\ref{Tisc} \& \ref{Tisa}), IS has approximately equal number of comparisons and assignments. This property makes BCIS runs faster than IS even when they have close number of comparisons.

\par Hence, we wrote the code section in statements (\ref{s13}-\ref{s20}) to optimize the performance of BCIS by keeping $ k$ close to $ n^{0.5}$ as possible. This code segment is based on the idea that ensures at least a set of length $(SR-SL)^{0.5} $ not to be inserted during the current sort trip (where sort trip size = SR-SL). This idea realized by scanning this set looking for the minimum and maximum element and replace them with LC and RC respectively. However, this code does not add extra cost for the performance of the algorithm because the current sort trip will start where the loop in statement (\ref{s14} ) has finished (sort trip will start at $ (SR-SL)^{0.5 }+1$). Theoretical results have been compared with experimental results in \hyperref [section:5]{section-\ref*{section:5}} and BCIS showed performance close to the best performance of average case  that explained above. 

\par In the rest part of this section, instruction level analysis of BCIS is presented. We re-analyze the algorithm for average case by applying above assumption to get more detailed analysis. However, the cost of each instruction is demonstrated as comment in the  pseudo code of the algorithm, we do not explicitly calculate the cost of the loop in statement (\ref{s14}), because it is implicitly calculated with the cost of not inserted elements inside the loop started in statement (\ref{s25}). Code segment within statements (\ref{s13}-\ref{s20}) activates for sort trip size greater than 100 elements only. Otherwise, sort trip index $i$  starts from the element next to SL (statement  \ref{s22}). 
\par The total number of comparisons for each insertion function is calculated by equation(\ref{compno}) multiplied by the number of sort trip $(n/k)$ as following:-
\begin{equation}
\frac{n}{k}\left(\frac{k^2}  {16}+ \frac{3k}{8}-1\right)= \frac{nk}  {16}+ \frac{3n}{8}- \frac{n}{k}
 \end{equation}
  The Complexity of the check equality function ISEQUAL is neglected because \textbf{if} statement at (\ref{s5}) rarely gets true. The total complexity of BCIS is calculated as following:-
 \begin{algorithm} 
\caption {BCIS Average case analysis  Part 1}
\begin{algorithmic}[1]
 \State $SL\leftarrow  left$  \Comment C1
  \State $SR\leftarrow  right$  \Comment C2
 \While {$SL<SR$} \Comment C3$(\frac{n}{k}+1)$
  \State SWAP$(array ,SR ,SL+\frac{(SR-SL)}{2}) $\Comment C4$(\frac{n}{k})$
\If {$array[SL] $= $array[SR]$}\Comment C5$(\frac{n}{k})$
\If{ISDUP($array,SL,SR$)=-1}  
\State return
\EndIf
\EndIf      
\If {array[SL] $>$ array[SR]} \Comment C6$(\frac{n}{k})$
\State SWAP (array, SL , SR)
\EndIf
\If {$ (SR-SL)\geq100$}\Comment C7$(\frac{n}{k})$
\For {i }{SL+1}{(SR-SL)^{0.5}}
\If {array[SR] $<$ array[i]}
\State  SWAP (array, SR, i)
\ElsIf  {array[SL] $>$ array[i]}
\State  SWAP (array, SL, i)
\EndIf
\EndFor
\Else 
\State  $ i\leftarrow SL+1$
\EndIf

\State $LC \gets array[SL]$ \Comment C8$(\frac{n}{k})$
\State $RC \gets array[SR]$\Comment C9$(\frac{n}{k})$
 \algstore{bkbreak} 
 \end{algorithmic}
 \end{algorithm}
 \begin{algorithm} [h]
 \caption{BCIS Average case analysis Part 2}
 \begin{algorithmic}[1]
 \algrestore{bkbreak}

 \While {$i<SR$} \Comment C10 $\sum_{i=1}^{\frac{n}{k}}(n-ik) $
\State $CurrItem \leftarrow array[i] $ \Comment { C11 $\sum_{i=1}^{\frac{n}{k}}(n-ik) $}
\If{$CurrItem \ge RC$} \Comment C12$\sum_{i=1}^{\frac{n}{k}}(n-ik) $
\State $array [i] \gets  array[SR-1] $ \Comment C13$(\frac{n}{2})$
\State $INSRIGHT (array, CurrItem, SR, right) $\Comment C14 $( \frac{nk}{16}+\frac{3n}{8}-\frac{n}{k})$
\State $SR \gets  SR-1 $                \Comment C15$(\frac{n}{2})$

\ElsIf{$CurrItem \le LC$} \Comment C16 $( \sum_{i=1}^{\frac{n}{k}}(n-ik) -\frac{n}{2} )$
\State $array [i] \gets  array[SL+1] $ \Comment C17$(\frac{n}{2})$
\State $INSLEFT (array, CurrItem, SL, left) $  \Comment C14 $( \frac{nk}{16}+\frac{3n}{8}-\frac{n}{k})$
\State $SL \gets  SL+1 $    \Comment C18$(\frac{n}{2})$
\State $i \gets i+1$    \Comment C19$(\frac{n}{2})$
\Else 
\State $i \gets i+1$ \Comment C20$(\sum_{i=1}^{\frac{n}{k}}(n-ik) -n )$
\EndIf
\EndWhile
\State $SL\leftarrow SL+1$    \Comment C19$(\frac{n}{k})$
\State $ SR\leftarrow SR-1$   \Comment C19$(\frac{n}{k})$
\EndWhile
 \end{algorithmic}
\end{algorithm} 
 
 \begin {align*} 
T(n) &=C1+C2+C3\\
       &+ (C3+C4+C5+C6+C7+C8+C21+C22)\frac{n}{k}\\
       &+(C10+C11+C12+C16+C20 ) \sum_{i=1}^{\frac{n}{k}} (n-ik)\\
       & +(C13+C15-C16+C17+C18+C19) \frac{n}{k}\\ 
       &+C14 *2 \left( \frac{nk}{16}+\frac{3n}{8}-\frac {n}{k}\right)- C20 n\\
  a&=(C3+C4+C5+C6+C7+C8+ C21+C22)\\
  b&=C14  \\                   
  c&=( C10+C11+C12+C16+C20)\\             
  d&=(C13+C15-C16+C17+C18+C19)\\      
  e&=C20     \\
  f&=(C1+C2+C3) \\
T(n)&=a\frac{n}{k}+b \left(  \frac{nk}{8}+\frac{3n}{4}-\frac {n}{k}  \right)+c \sum_{i=1}^{\frac{n}{k}} (n-ik)\notag \\
        &+d \frac{n}{2}- en + f\notag \\
        &=a\frac{n}{k}+b \left(  \frac{nk}{8}+\frac{3n}{4}-\frac {n}{k}  \right)+c \left( \frac{n^2}{2k}-\frac{n}{2} \right)\notag\\
        &+d \frac{n}{2}- en + f\notag \\
\end{align*}

\begin{align}
        &=\frac{n}{k}\left[a+b \left(  \frac{k}{8}+\frac{3k}{4}-2  \right)+c\frac{n}{2} \right]-c\frac{n}{2}\notag\\
        &+d \frac{n}{2}- en + f \label{TnIns}
\end{align}
\par We notice that equation (\ref{TnIns}) is similar to equation (\ref{Tc}) when constants represent instructions cost.

\subsection{Best Case Analysis}
\par
\begin{figure}[h]
\includegraphics [width=1\linewidth]{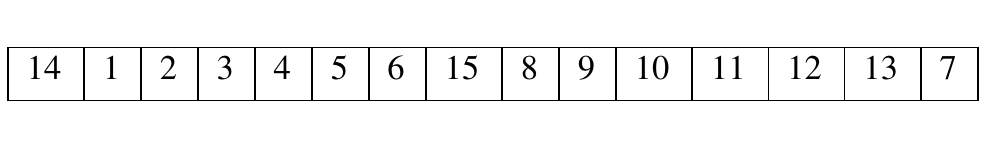}
\caption{ Best case example for array less than 100 elements }
    \label{fig:BestCase1}
\end{figure}

 The best case occurs in case of every element  is placed in its final correct position consuming a limited and constant number of comparisons and shift operations at one sort trip. These conditions are available  once the first sort trip starts while the RC and LC are holding the largest and second largest item in the array respectively, and all other elements are already sorted. The following example in Figure(\ref{fig:BestCase1}) explains this best case (the element 15 will replaced with 7 by the statement \ref{s4}).

 For this  best case, we note that all  insertions will be in the left side only with one shifting operation per each insertion. That means the cost of insertion each item is $O(1)$. Therefor, the total cost of the left insertion function is $T_{is} (n)=n$. Also all elements will inserted in one sort trip so that $k=n$ .These values is substituted in equation(\ref{Eq1}) as follows:-\\
\begin{align}
 T(n)=&\frac{n}{k} \left[ T_{is} (n) \right]+\sum _{i=1}^{\frac{n}{k}}(n-ik)\notag\\
         &where \>  k=n \notag \\
 T(n)=&n \label{eqBest1}
 \end{align}
Hence, the best case of BCIS is O (n) for $n <100$. Otherwise, (for $n\geq 100$) the loop started in statement (\ref{s14}) always prevents this best case occurred because it only put LC and RC in their correct position  and disallow insertions during all sort trips. As result, the loop in statement (\ref{s14}) forces the algorithm running very slow on already sorted or revers sorted array.

\par Generally, already sorted and reverse sorted arrays are more common in practice if compared with the above best case example. Therefore, statement (\ref{s4}) has been added to enhance the performance of best case and worst case when BCIS run on sorted and revers sorted arrays. In case of already sorted array, this statement makes the BCIS, during each sort trip, inserts half of (SR-SL) in least cost.
\par
\begin{figure}[h]
   \centering
   \includegraphics [width=1\linewidth]{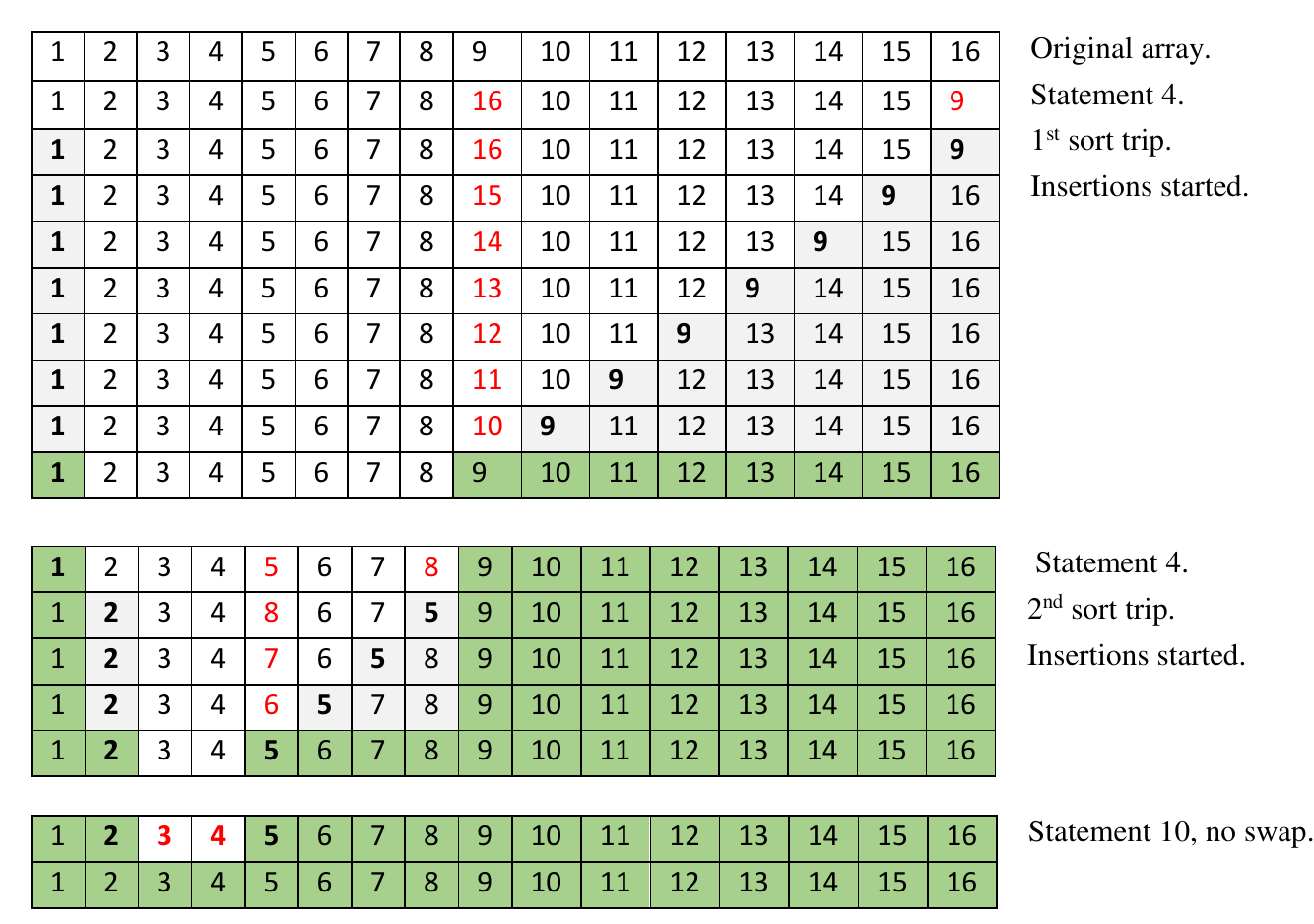} 
    \caption{Example of running BCIS on already sorted array}
    \label{fig:BestCase2}
\end{figure}

 The following example in Figure (\ref{fig:BestCase2}) explains how BCIS runs on already sorted array. For simplicity not inserted elements are not represented in each sort trip during the demonstration of this example.

\par For already sorted array, BCIS scans the first half consuming two comparisons per item (no insertions), then inserts the second half of each sort trip consuming two comparisons per item too. Because the sort trip size is repeatedly halved. Hence, it can represent as following.
\begin{align}
T(n)=&\left (2 \frac{n}{2}+2 \frac{n}{2}\right)+\left (2 \frac{n}{4}+2 \frac{n}{4}\right)\notag\\
          &+\left (2 \frac{n}{8}+2 \frac{n}{8}\right) +.....+\left (2 \frac{n}{2^i}+2 \frac{n}{2^i}\right)\notag \\      
&stop\>when \>2^i=n \implies  i=log\> ⁡n \notag\\
T(n)=&\sum_{i=1} ^{log\>⁡n}(2 \frac{n}{2^i}+2 \frac{n}{2^i}) =4n \label{eqBest2}
\end{align}
\par Equations (\ref{eqBest1} \& \ref{eqBest2})  represent the best case growth functions of BCIS when run on array size less than 100 and greater than  and equal to 100 respectively.

\subsection{Worst Case Analysis}
\begin{figure}[h]
    \centering
         \includegraphics [width=1\linewidth]{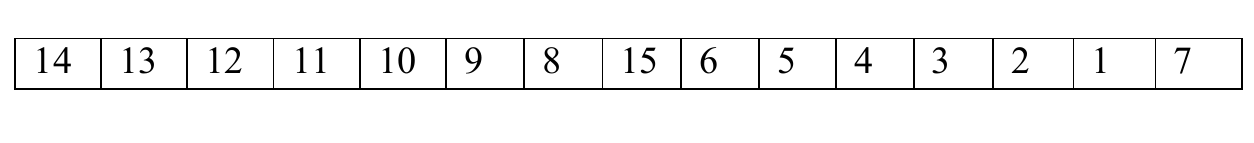} 
    \caption{ Worst case example for array less than 100 elements}
    \label{fig:WorstCase1}
\end{figure}

The worst case happens only if all elements are inserted in one side in reverse manner during the first sort trip. This condition provided when the RC and LC are the largest and second largest numbers in the array respectively, and  all other items are sorted in reverse order. The insertion will be in the left side only. The following example in Figure(\ref{fig:WorstCase1}) explains this worst case when $n < 100$.
\par Since each element in the this example inserted reversely, the complexity of left  insertion function for each sort trip equal to $T_{is} (n)=\frac{n(n-1)}{2}$. Also there is one sort trip so $k=n$, by substitute these values in equation(\ref{Eq1}) as follows :-
  \begin{align}
 T(n)=&\frac{n}{k} \left[ T_{is} (n) \right]+\sum _{i=1}^{\frac{n}{k}}(n-ik)\notag\\
         &where \>  k=n \notag \\
 T(n)=& \frac{n(n-1)}{2} \label{eqWorst1}
 \end{align} 
\par Hence, the worst case of BCIS is $ O (n^2)$ for  $n < 100$ . Likewise the situation in the best case, the loop in statement (\ref{s14}) prevent the worst case happen because LC will not take the second largest item in the array. Consequently, the worst case of  BCIS  would be when it runs on reversely sorted array for $ n\geq 100$. The following example explains the behaver of the BCIS on such arrays even the size of array  less than 100.

\begin{figure}[h]
    \centering
         \includegraphics [width=1\linewidth]{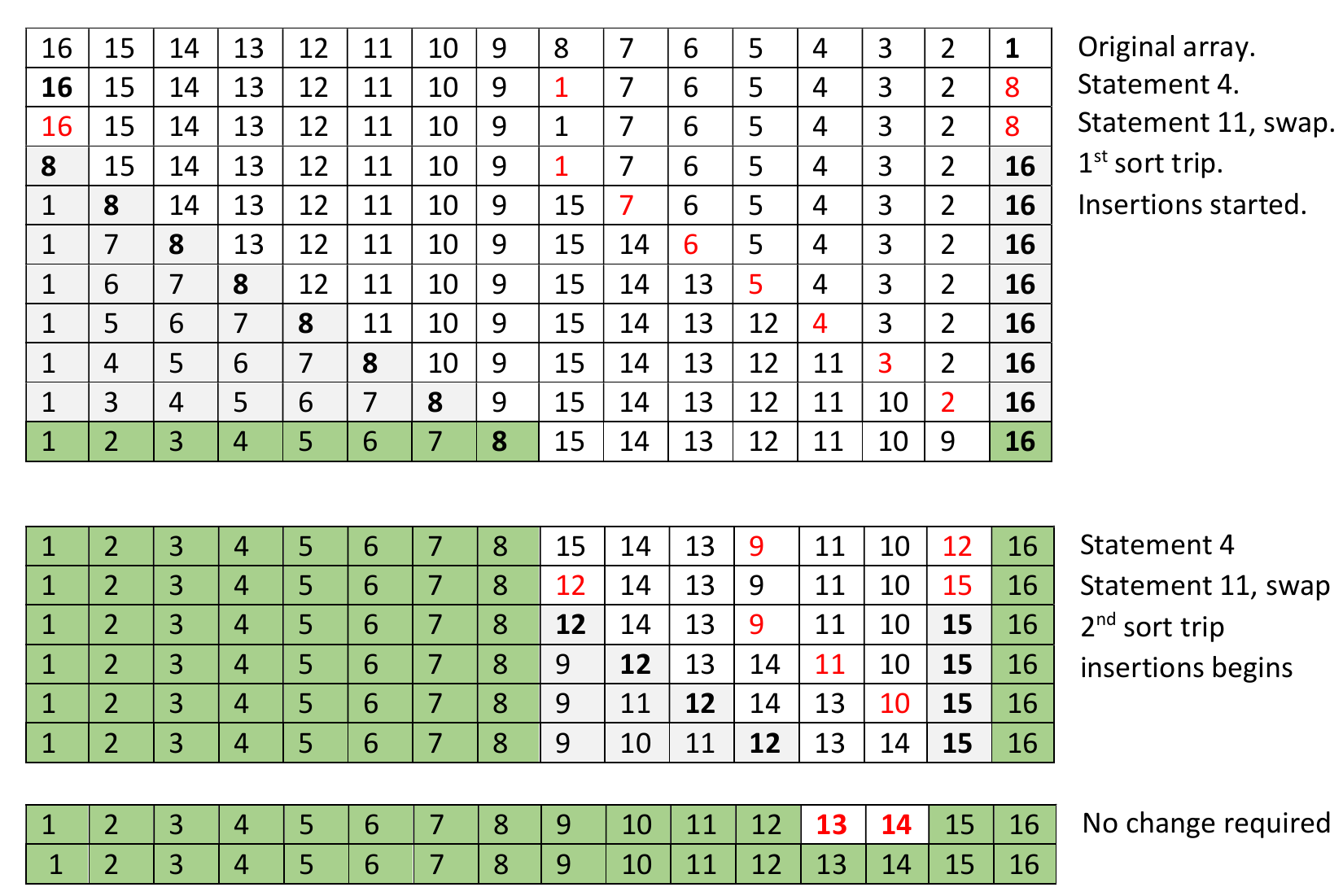}
         \caption{ Example of running BCIS on reversely sorted array}
    \label{fig:WorstCase2}
\end{figure}
\par In case of reversely sorted array, BCIS does not insert the first half of the scanned elements, cost two comparisons per each element, then insert the second half reversely for each sort trip approximately. Considering the cost of reverse insertion (for each sort trip) is $T_{is}(k)=\frac{k(k-1)}{2}$ where $k$ halved repeatedly. Like already sorted array analysis, the complexity of BCIS can be represented as follows. 
 \begin{align}
T(n)=&\left (2 \frac{n}{2}+ \frac{(\frac{n}{2})^2 - \frac {n}{2}}{2}\right) +\left (2 \frac{n}{4}+ \frac{(\frac{n}{4})^2 - \frac {n}{4}}{2}\right) \notag\\   
+&\left (2 \frac{n}{8}+ \frac{(\frac{n}{8})^2 - \frac {n}{8}}{2}\right)  +........\notag\\   
+&\left (2 \frac{n}{2^i}+ \frac{(\frac{n}{2^i})^2 - \frac {n}{2^i}}{2}\right)\notag\\ 
&stop\>when \>2^i=n \implies  i=log\> ⁡n \notag\\
T(n)=&\sum_{i=1} ^{i=log\>⁡n}\left (2 \frac{n}{2^i}+ \frac{(\frac{n}{2^i})^2 - \frac {n}{2^i}}{2}\right) \notag\\ 
       =&\frac{n^2}{6}+\frac{3n}{2} \label{eqWorst2}
\end{align}
\par Equations (\ref{eqWorst1} \& \ref{eqWorst2})  represent the worst case growth functions of BCIS when run on array size less than 100 and greater than  and equal to 100 respectively.

\section {Results and comparison with other algorithms}\label{section:5}
The proposed algorithm  is implemented by C++ using NetBeans 8.0.2 IDE based on Cygwin compiler. The measurements are taken on a 2.1 GHz Intel Core i7   processor with 6 GB 1067 MHz DDR3 memory machine with windows platform. Experimental test has been done on an empirical data (integer numbers) that generated randomly using a C++ class called “uniform\_int\_distribution”. This class  generates specified ranged of random integer numbers  with uniform distribution\cite{CplusplusUniform}.
\subsection{BCIS and classical insertion sort}
\begin{figure}[h]
    \centering
         \includegraphics [width=1\linewidth]{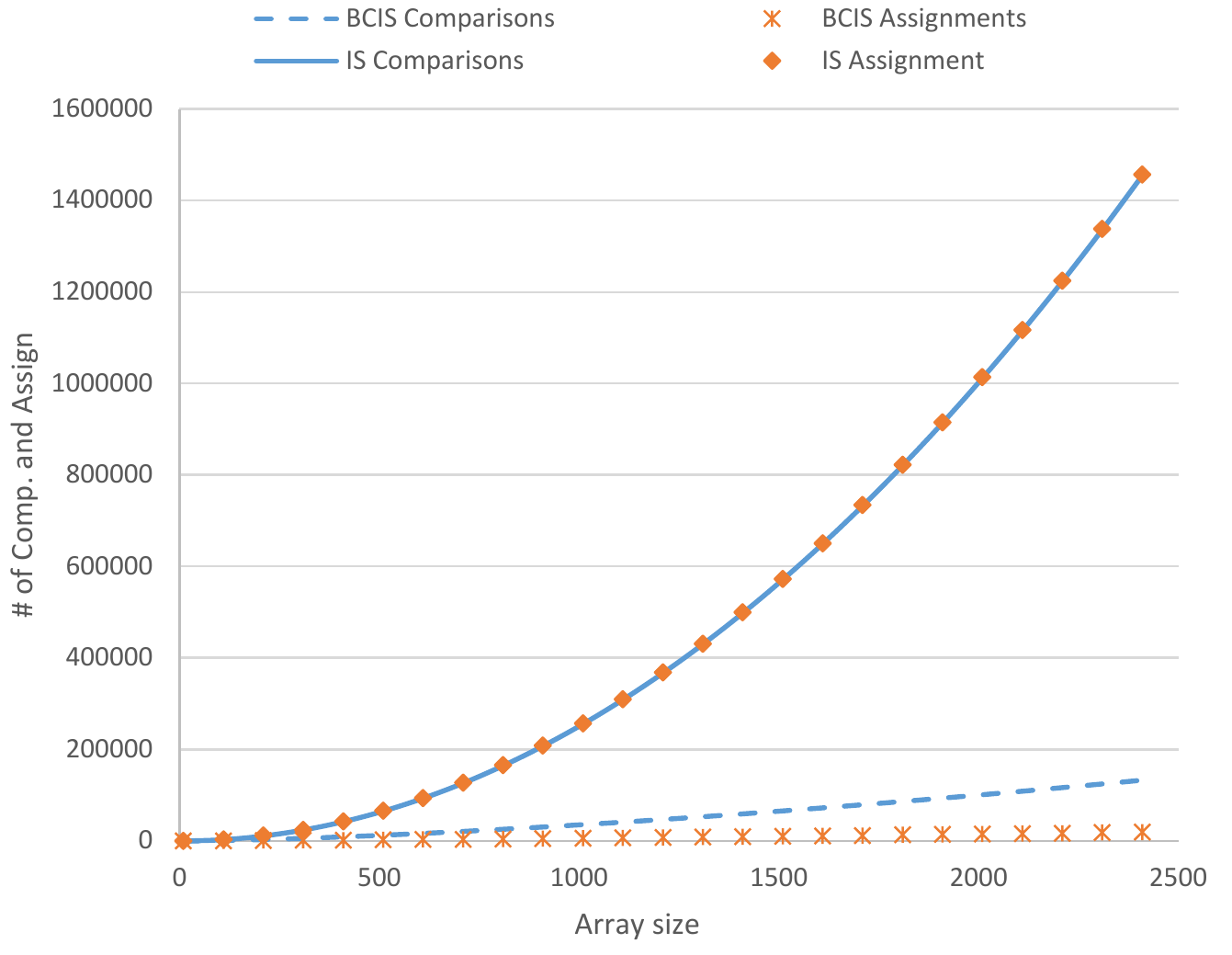}
 
    \caption{ No. of comparisons and assignments  for  BCIS and IS}
    \label{fig:results1}
\end{figure}

Figure \ref{fig:results1} explains the average number of comparisons and assignments (Y axis) for BCIS and IS with different list size (X axis). The figure has been plotted using equations (\ref{Tisc},\ref{Tisa},\ref{TcBCIS} \& \ref{TaBCIS}).  This figure explains that the number of comparisons and assignments of IS  are approximately equal . In contrast, in  BCIS the number of assignments are less than the number of comparisons. This feature show the better performance of BCIS and support our claim that BCIS has less memory read/write operations when compared with IS.
\par Though the theoretical average analysis for BCIS and IS is calculated in term of the number of comparisons and assignments separately in equations (\ref{Tisc},\ref{Tisa},\ref{TcBCIS} \& \ref{TaBCIS}). In order to compare the results of these equations with experimental results of BCIS and IS which are measured by execution elapsed time, we represent these quantities as a theoretical and experimental ratio of $\frac {BCIS}{IS}$. In theoretical ratio we assumed that the cost of an operation of  comparison and assignment is equal in both algorithms. Therefor, equation (\ref{Tisc}) has been added to equation(\ref{Tisa}) to compute the total cost of IS ($IS_{total}$). Similarly, the total cost of BCIS ($BCIS_{total}$) is the result of add equation (\ref{TcBCIS}) to equation (\ref{TaBCIS}).

\par
\begin{figure}[h]
    \centering
         \includegraphics [width=1\linewidth]{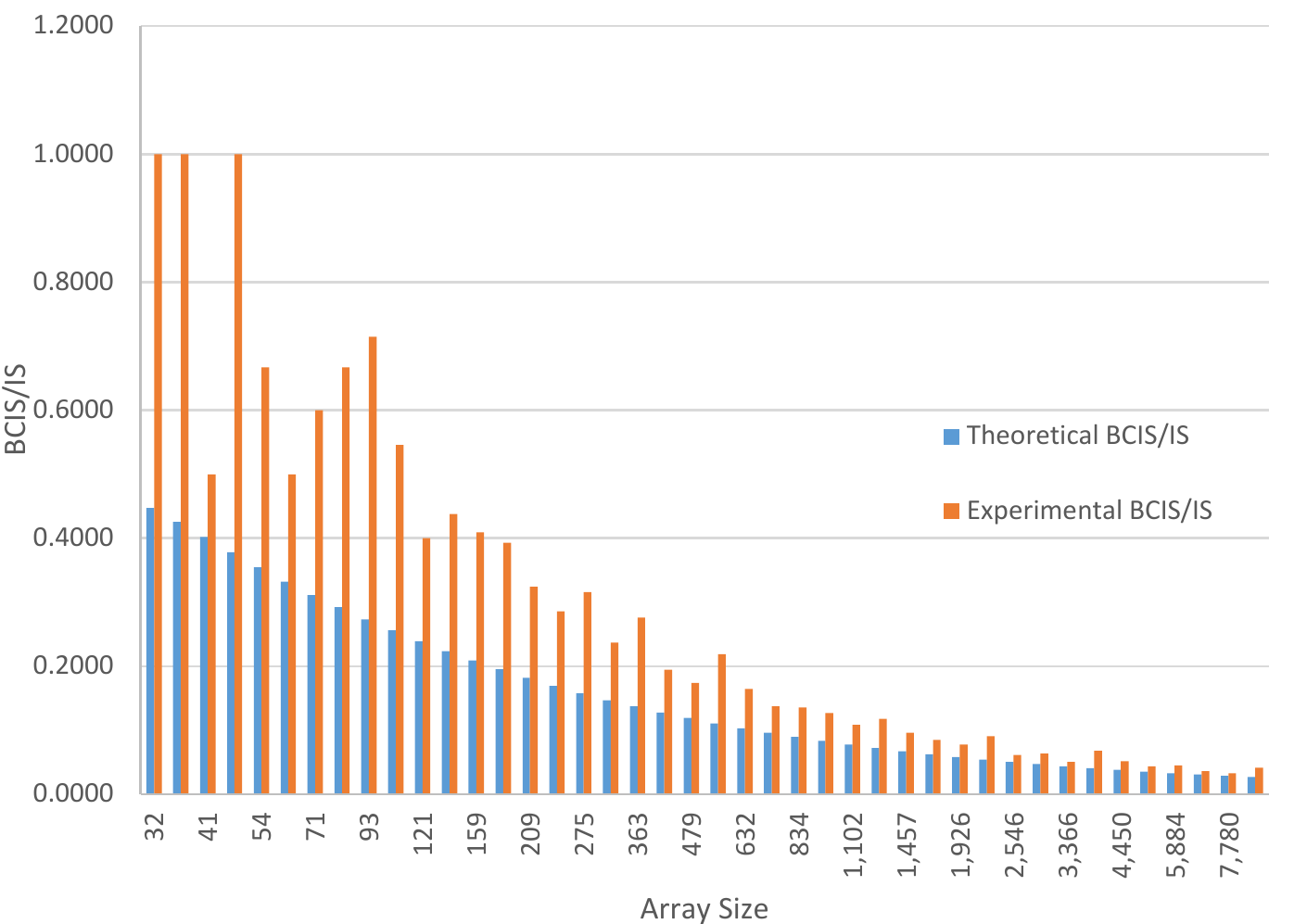}
 
    \caption{BCIS/IS ratio for $ n<10000$}
    \label{fig:results2}
\end{figure}

 Figure(\ref{fig:results2}) illustrates a comparison in performance of the proposed algorithm BCIS and IS.  This comparison has been represented in terms of the ratio  BCIS/IS  (Y axis) that required to sort a list of random data for some list sizes (X axis). Theoretically, this ratio is equal to($\frac{BCIS_{total}}{IS_{total}}$) . In opposition, the experimental ratio computed by divide $BCIS_{time}$ over $IS_{time}$, when these parameters represent experimental elapsed running time of BCIS and IS respectively.

\par
\begin{figure}[h]
    \centering
         \includegraphics [width=1\linewidth]{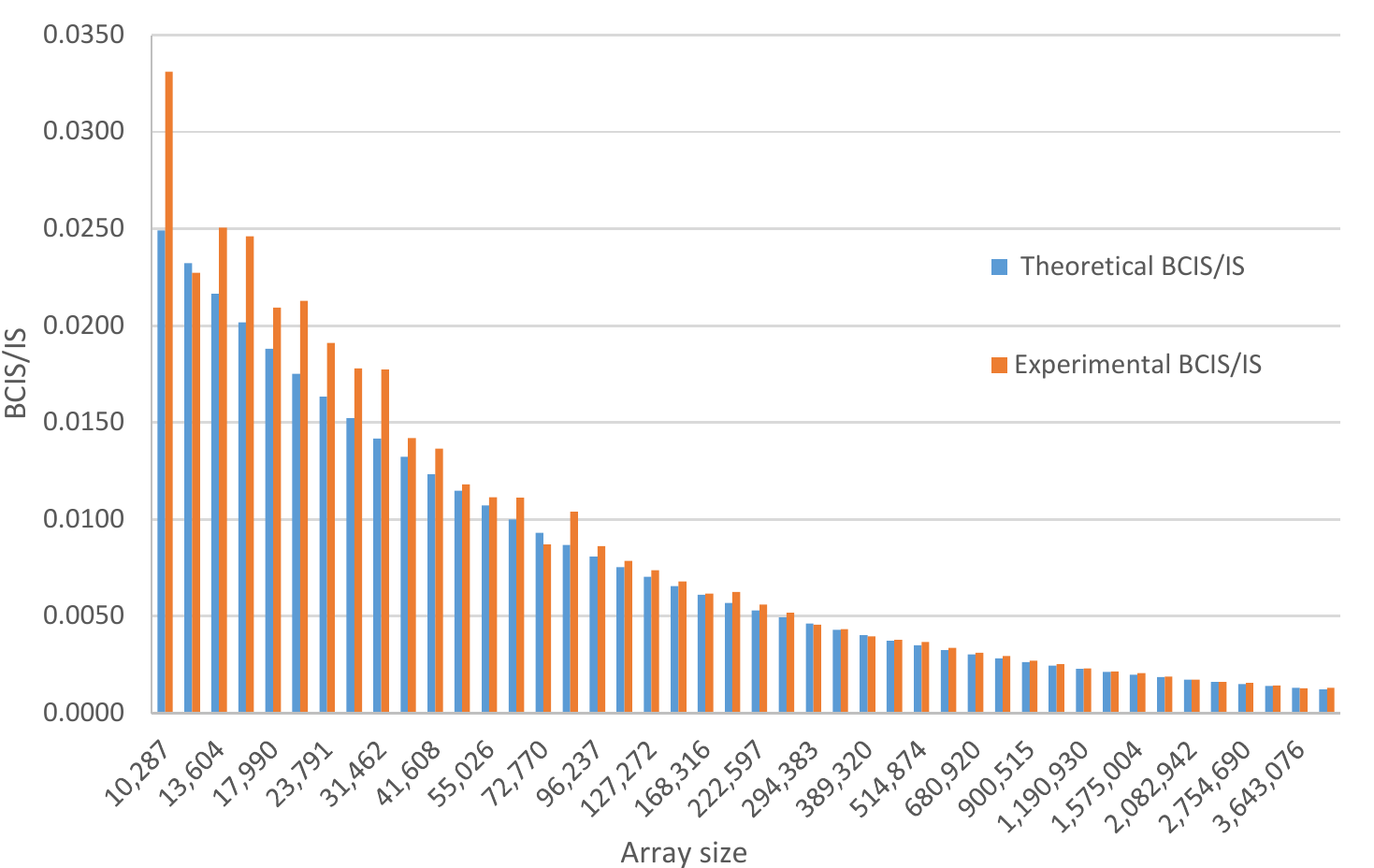}
 
    \caption{BCIS/IS ratio for $n>10000$}
    \label{fig:results3}
\end{figure}

 In the experimental $\frac{BCIS _{time}}{IS_{time}}$   ratio, we noticed that the proposed algorithm has roughly equal performance when compared to classical insertion sort for list size less than 50 ($\frac{BCIS _{time}}{IS_{time}} =1$). However, the performance of BCIS increased for larger list size noticeably. The time required to sort the same size of list using BCIS begin in 70\% then inclined to 4\% of that consumed by classical insertion sort for list size up to 10000. Figure (\ref{fig:results3}) explains the same ratio for n>10000. This figure shows that  $\frac{BCIS _time}{IS_time}$  is decreased when the list size increased. For instance for the size 3,643,076 the experimental ratio equal to 0.00128 that means BCIS 781 times faster than IS. 
\par In conclusion, Figures (\ref{fig:results2}\& \ref{fig:results3})  show that the theoretical an experimental ratio are very close especially for large size lists. This means BCIS go close to the best performance of average case for large size lists.

\subsection{BCIS and Quicksort}
\subsubsection{BCIS and Quicksort comparison for no duplicated-elements data set}
\begin{figure}[h]
   \centering
    \includegraphics [width=1\linewidth]{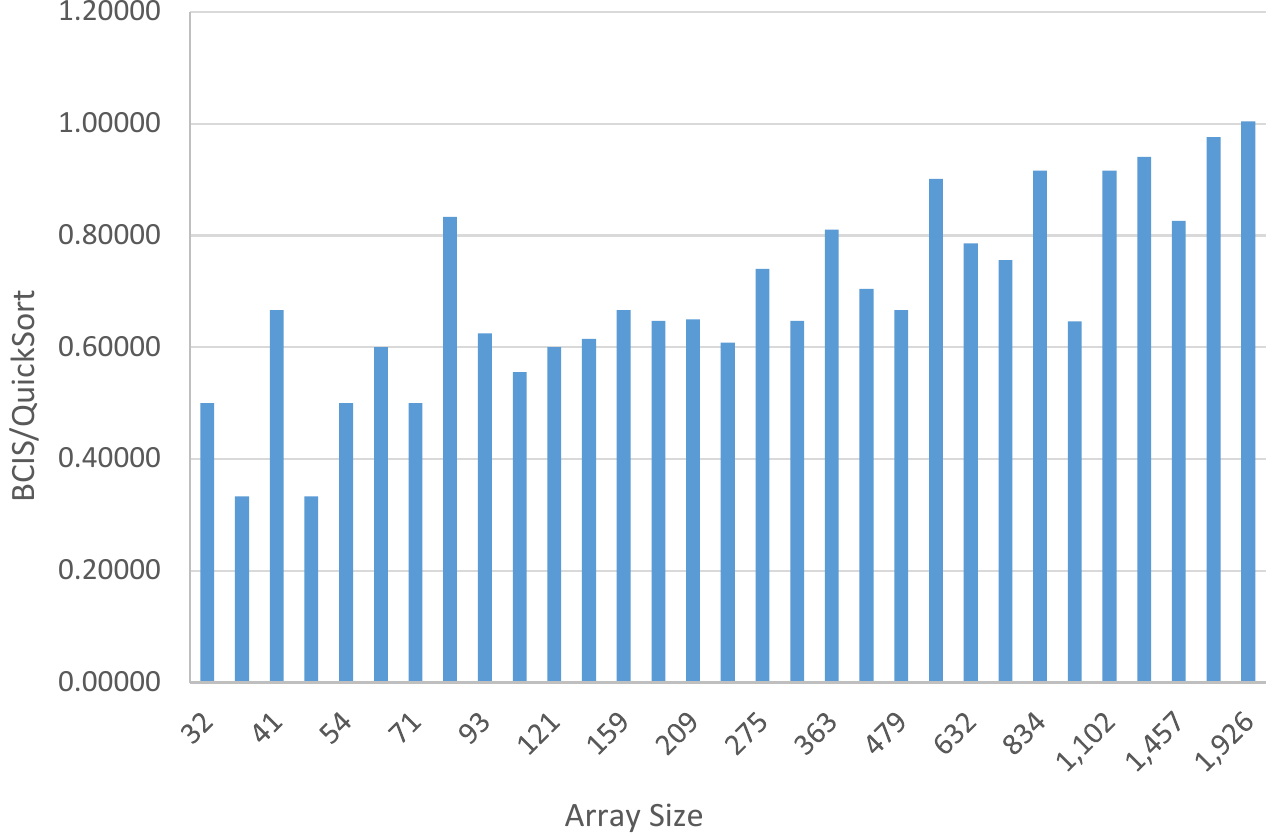} 
    \caption{BCIS and Quick Sort performance $n<2000$}
    \label{fig:results4}
\end{figure}

Figure (\ref{fig:results4}) explains a comparison in experimental performance of the proposed algorithm BCIS and Quicksort for small size lists. Widely used enhanced version of Quicksort (median-of-three pivots) is used, which is proposed by \cite{Sedgewick1977}. This comparison has been represented in terms of  the experimental ratio  $\frac{BCIS _{time}}{QuickSort_{time}}$ (Y axis) that required to sort a list with random data for some list sizes (X axis). We remarked that BCIS is faster than Quicksort for list size less than 1500 for most cases. The time required to sort the same size of list using BCIS ranged between 30\% and 90\% of that consumed by Quicksort when list size less than 1500.
\par
\begin{figure}[h]
   \centering
    \includegraphics [width=1\linewidth]{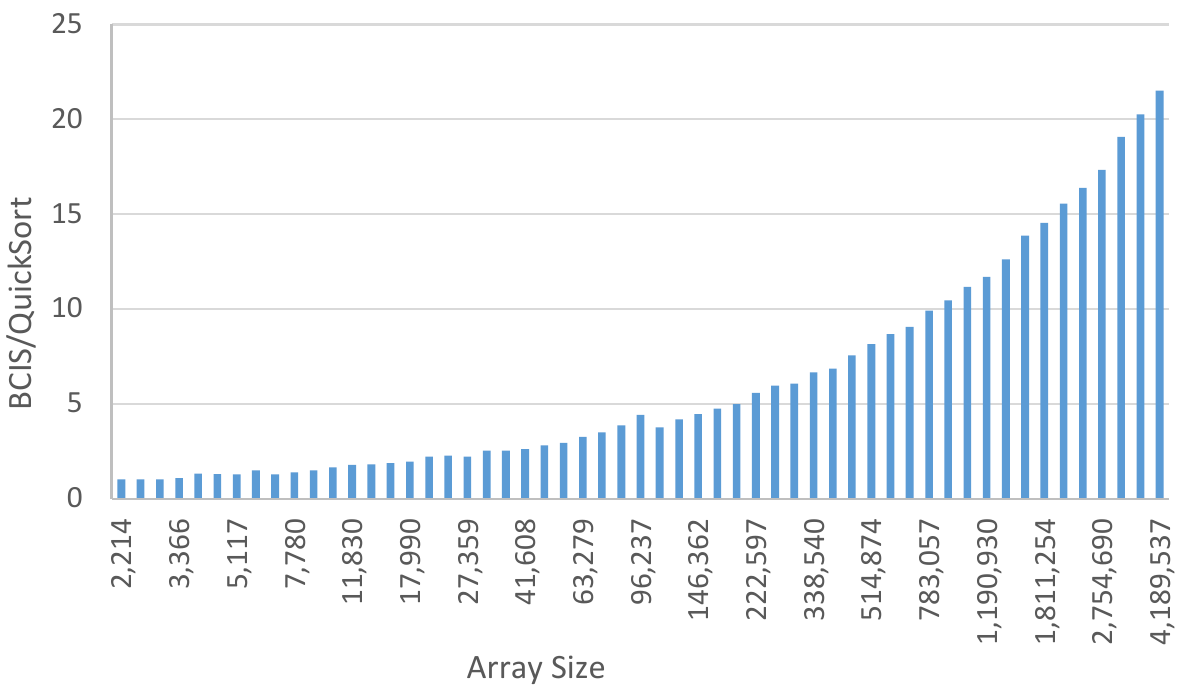}
     \caption{BCIS/Quicksort for $ n>2000$}
    \label{fig:results5}
\end{figure}

 Although  theoretical analysis in all previous cited works that have been explained in literature (\hyperref [section:1]{section-\ref*{section:1}}) show that  QuickSort has more efficient comparison complexity than BCIS. But experimentally  BCIS defeats QuickSort for relatively small array size for some reasons. First the low number of assignment operations in BCIS, second  an assignment operation has lower cost if compared with swap operation that used in QuickSort, whereas each swap operation requires three assignments to done. Finally, due to the nature of the cache memory architecture\cite{handy1998cache}, BCIS uses cache memory efficiently because shifting operations only  access to the adjacent memory locations while swap operations in QuickSort access memory randomly. Therefore, QuickSort cannot efficiently invest the remarkable speed gain that is provided by the cache memory. 
However, Figure (\ref{fig:results5}) shows  experimental  $\frac{BCIS _{time}}{QuickSort_{time}}$ ratio for array size greater than 2000 and less than 4,500,000. 
\subsubsection{BCIS and Quicksort comparison for high rate of duplicated elements data set}
\begin{figure}[h]
    \centering
    \includegraphics [width=1\linewidth]{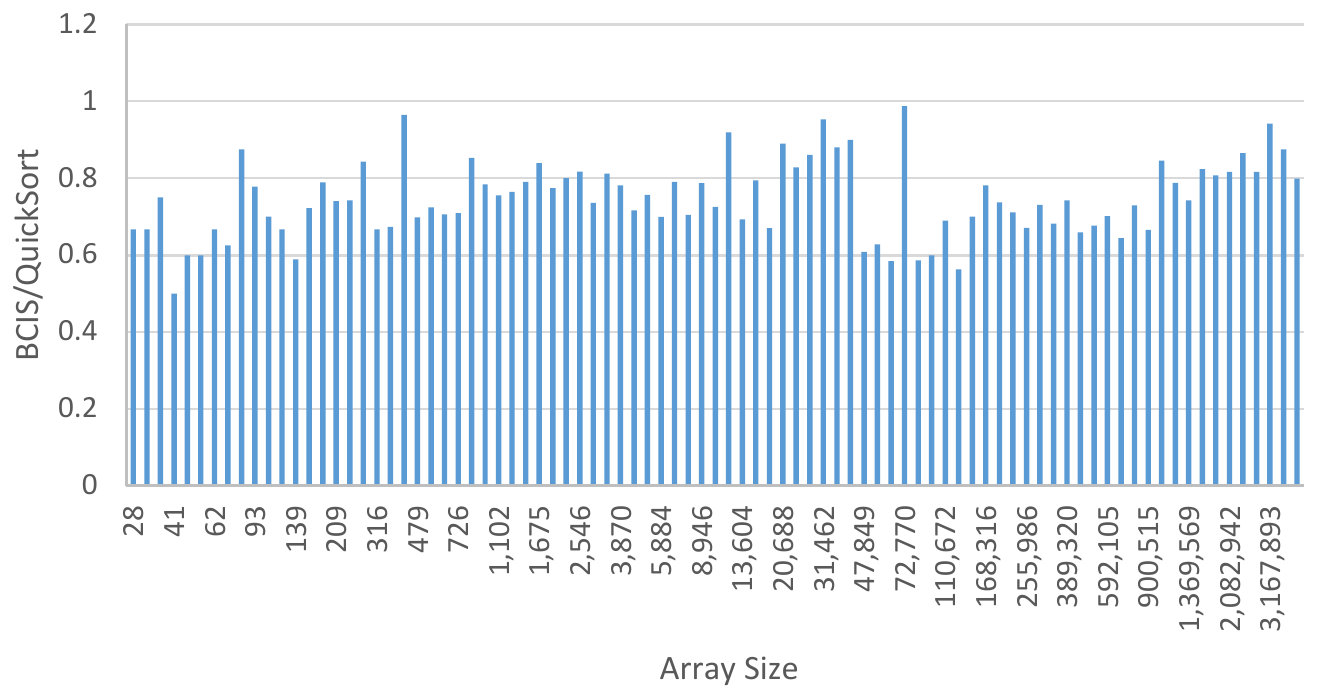} 
    \caption{experimental BCIS/Quicksort ratio for 50 duplicated elements}
    \label{fig:results6}
  \end{figure}

Experimental test showed that BCIS faster than Quicksort when running on data set has high rate of duplicated elements even for large list size. Figure (\ref{fig:results6}) explains the experimental  ratio $\frac{BCIS}{Quicksort}$  when the used array has only 50 different elements. The computer randomly duplicates the same 50 elements for arrays that have size greater than 50. This figure shows that BCIS consumes only 50\% to 90\% of the time consumed by Quicksort when run on high rate of duplicated elements array. This variation in ratio is due to the random selection of LC and RC during each sort trip. The main factor that make BCIS faster than Quicksort for such type of array is that there a small number of assignments and comparisons operations if there are many numbers equal to LC or RC in each sort trip. This case could occur with high probability when there is high rate of duplicated elements in array.  
   \section{Conclusions}\label{section:6}
 In this paper we have proposed a new bidirectional conditional insertion sort. The performance of the proposed algorithm has significant enhancement over classical insertion sort.  As above shown results prove. BCIS has average case about  $n^{1.5 }$  . Also BCIS keeps the fast performance of the best case of classical insertion sort when runs on already sorted array. Since BCIS time complexity is (4n) over that array. Moreover, the worst case of BCIS is better than IS whereas BCIS consumes only $ \frac{n^2}{6}$  comparisons with reverse sorted array.
\par The other advantage of BCIS is that algorithm is faster than Quicksort for relatively small size arrays (up to 1500). This feature does not make BCIS the best solution for relatively small size arrays only. But it makes BCIS powerful interested algorithm to use in conjugate with quick sort. The performance of the sorting process for large size array could be increased using hybrid algorithms approach by using Quicksort and BCIS. Additionally, above results shown that BCIS is faster than quick sort for arrays that have high rate of duplicated elements even for large size arrays. Moreover, for fully duplicated elements array, BCIS indicates fast performance as it can sort such array in only O(n).

\section *{Acknowledgments}
This research is partially supported by administration of governorate of Salahaddin and Tikrit university - Iraq. We thank Thamer Fahad Al-Mashhadani  for comments that greatly improved the manuscript.

\begin{wrapfigure}{l}{20mm} 
    \includegraphics[width=1in,height=1.25in,clip,keepaspectratio]{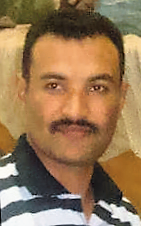}
  \end{wrapfigure}
\textbf{Adnan Saher Mohammed }  received B.Sc degree in 1999 in computer engineering technology from College of Technology, Mosul, Iraq. In 2012 he obtained M.Sc degree in communication and computer network engineering from UNITEN University, Kuala Lampur, Malasyia. He is currently a Ph.D student at graduate school of natural sciences,Y{\i}ld{\i}r{\i}m Beyaz{\i}t University, Ankara, Turkey. His research interests include Computer Network and computer algorithms.\\ \\
\par
\begin{wrapfigure}{l}{20mm} 
    \includegraphics[width=1in,height=1.25in,clip,keepaspectratio]{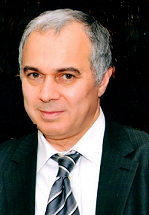}
  \end{wrapfigure}\par
  \textbf{\c{S}ahin Emrah Amrahov} received B.Sc. and Ph.D. degrees in applied mathematics in 1984 and 1989, respectively, from Lomonosov Moscow State University, Russia. He works as Associate Professor at Computer Engineering department, Ankara University, Ankara, Turkey. His research interests include the areas of mathematical modeling, algorithms, artificial intelligence, fuzzy sets and systems, optimal control, theory of stability and numerical methods in differential equations.\\
\par
\begin{wrapfigure}{l}{20mm} 
    \includegraphics[width=1in,height=1.25in,clip,keepaspectratio]{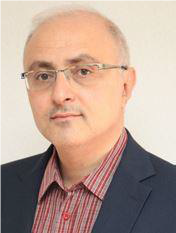}
  \end{wrapfigure}\par

\textbf{Fatih Vehbi \c{C}elebi}  obtained his B.Sc degree in electrical and electronics engineering in 1988, M.Sc degree in electrical and electronics engineering in 1996, and Ph.D  degree in electronics engineering in 2002 from Middle East Technical University, Gaziantep University and Erciyes University respectively. He is currently head of the Computer Engineering department and vice president of Y{\i}ld{\i}r{\i}m Beyaz{\i}t University,Ankara-Turkey. His research interests include Semiconductor Lasers, Automatic Control, Algorithms and Artificial Intelligence.

\end{document}